\documentstyle[preprint,twoside,adbis962]{acmconf}
\author{L.Yu. Ismailova, K.E. Zinchenko \vspace{1.52mm} \\
{ Vorotnikovskiy per, 7, bld. 4} \\
{ Institute for Contemporary Education ``JurInfoR-MSU ''} \\
{ Moscow, 103006, Russia}\\
{\tt larisa@jurinfor.ru}}
\addtocounter{footnote}{1}
\title{\bf Object-oriented tools for advanced applications \\%
(an extended abstract)
}

\begin{document}

\bibliographystyle{alpha}

\maketitle


\begin{abstract}
{\small This paper contains a brief discussion of the Application
Development Environment (ADE) that is used to build database
applications involving the graphical user interface (GUI). ADE
computing separates the database access and the user interface.
The variety of applications may be generated that communicate with
different and distinct desktop databases. The advanced techniques
allows to involve remote or stored procedures retrieval and call.
}
\end{abstract}


\section*{Introduction}

Recent research activity generated not only the valuable advance in
understanding the nature of object but stimulated the experimental efforts in
development the object-oriented tools.

Here is briefly discussed the Application Development Environment
(ADE) that is used to build database applications involving the
graphical user interface (GUI). ADE computing separates the
database access and the user interface. The variety of
applications may be generated that communicate with different and
distinct desktop databases. The advanced techniques allows to
involve remote or stored procedures retrieval and call.

According to an object-oriented traditions \cite{OMG1:91},
ADE include some basic features
of inheritance, encapsulation, and polymorphism. They are used to derive an
actual object to cover the needed information resources.

The {\em potential object} (PO) is composed with the {\em menu} (M),
{\em data access} (DA),
and {\em modular counterparts} (MC).
The {\em Ancestor Potential Objects} (APO) contain
the menus, events, event evolver, attributes and functions (that are
encapsulated). The {\em Descendant Potential Objects} (DPO) are inherited from
APO.

The aim of the current contribution is to give a brief profile of
the ADE project without any detailed mathematical or
implementational consideration. Nevertheless, some mathematical
background corresponds to the references \cite{OMG1:91},
\cite{HeSa:95}.
Other less traditional
for the database area ideas are due to \cite{Wo:96}
to conform the {\em object computation} strategies. The main ADE
building blocks have the relative uniformity to resolve
the modular linkages. ADE enables the host computational
environment to extend the properties of the distinct MC.

\section{Event driven objects}

\begin{figure}
\footnotesize
\unitlength=1.00mm \special{em:linewidth 0.4pt}
\linethickness{0.4pt}
\begin{picture}(52.00,60.00)
\put(23.00,53.00){\makebox(0,0)[cc]{{\bf P}otential {\bf
O}bjects}} \put(21.00,15.00){\framebox(10.00,32.00)[cc]{}}
\put(26.00,43.00){\circle{2.00}} \put(26.00,33.00){\circle{2.00}}
\put(26.00,23.00){\circle{2.00}}
\put(23.00,36.00){\makebox(0,0)[cc]{h}}
\put(3.00,39.00){\makebox(0,0)[cc]{{\bf E}vents}}
\put(1.00,1.00){\framebox(10.00,32.00)[cc]{}}
\put(6.00,29.00){\circle{2.00}} \put(6.00,19.00){\circle{2.00}}
\put(6.00,9.00){\circle{2.00}}
\put(3.00,22.00){\makebox(0,0)[cc]{i}}
\put(7.00,20.00){\vector(2,1){40.00}}
\put(44.00,60.00){\makebox(0,0)[cc]{{\bf A}ctual {\bf O}bjects}}
\put(42.00,22.00){\framebox(10.00,32.00)[cc]{}}
\put(47.00,50.00){\circle{2.00}} \put(47.00,40.00){\circle{2.00}}
\put(47.00,30.00){\circle{2.00}}
\put(46.00,43.00){\makebox(0,0)[cc]{h(i)}}
\end{picture}
\caption{ {\bf E}vent {\bf D}riven {\bf O}bjects. {\small\em Here:
possible object $h$ is the mapping from the event (assignment) $i$
into the actual object $h(i)$. Note that a set of the possible
objects $\{h|h:I \to T\}=H_T(I)$ represents an idea of
functor-as-object for $I$ is a category of events, T is a
(sub)category of the actual objects - type. } }\label{pic:A}
\end{figure}
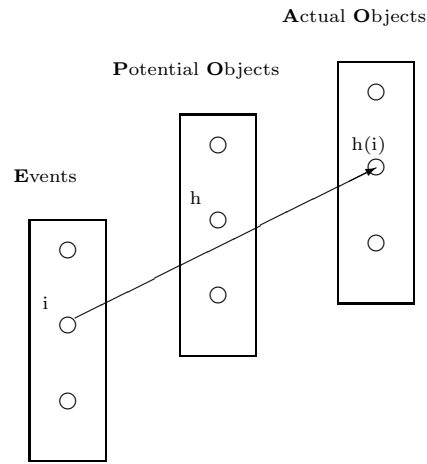

The MC is a holder of all the controls to communicate with the
user. The event is assigned by the user call (for instance,
clicking) or selection. Thus, when the activity is initiated, the
following main events may be triggered: respond to a request from
the user application, database retrieval or updating. The possible
order of the events is prescribed by {\em evolver} and is
determined by the {\em scripts}. A fragment of the event driven
procedure is shown in Fig.~\ref{pic:A}

\subsection{Menu}

Menu gives more flexibility to the attribute selection. Usually
the lists of possible attributes are supported to give the
developer or user more freedom. Menus are established to be
encapsulated in APO and are inherited in DPO.

\subsection{Particular application}

The particular application is derived from Potential Object Library (POL)
giving rise to Actual Object Libraries (AOL).

\section{Computational and relative backgrounds}

ADE concepts are based on a variant of computation theory for
information systems. The object-oriented programming, data base
and knowledge base engineering are included. Some vital concepts
have clear mathematical representations: {\em data object} that
represents the computer stored data; {\em metadata object} that
represents the conceptual information; {\em assignment} that
captures more dynamics and intrinsic states; {\em expansible
database} that represent the individualized self contained object
couples. The main feature of the general approach is in embedding
the typed entities (also the variable concepts) into the typeless
system that is based on the a theory of the variable concepts. The
last one is fitted to capture the dynamics of different objects
and switching the states in the information systems. The
prescribed modes in the information systems as a rule are
corresponded to the known classic cases: the relational database
theory, frame theory and logical reasoning in knowledge systems,
theory of programming.
     The implementation results in a non-Coddian notion of (relational)
database management system. Coddian type of database is equipped
with the first order relational logic but the current data
objects' base needs a higher order logic with the {\em
descriptions}. To extract its possible advantages the unified
architecture based on extensible data objects' model is proposed.
It supports the main object-oriented mechanisms of encapsulation,
polymorphism and inheritance and contains five major components.
They are Conceptual Shell (CS), Application Development
Environment (ADE), Basic Relational Tool System (BRTS), and
MetaRelational Tool System (MRTS). Some general counterparts
neatly corresponding to ADE are reflected in Fig.~\ref{pic:A}.
These subsystems are relatively self-contained, and  incorporated
into the ToolKit. BRTS supports relational interfaces ({\em
extensional level}), MRTS adds the {\em intentional level}, and
the pair $<$BRTS,~MRTS$>$ gives the enhanced relational features.
ADE contains the extensible data objects' model and manipulates
the switching, or {\em variable concepts} (references may be found
in~\cite{Wo:96}). Both BRTS and MRTS are embedded into ADE. CS
adds an external interactive mode to maintain and generate
applications.

\section{The relative approaches}

The ADE notion is based on the theory of variable concepts in the form of
variable sets originate from F.Lawvere investigations carried out in 1970's.
These constructions were used in elementary category theory, in particular in
representation of predicate logic. Until present time these results were of
interest only for theoreticians. In computer science variable sets are used in
the form of variable domains' theories in denotational semantics of
programming languages (pioneer investigations by C.Starchey \&
C.Wadsworth [1974]).

Variable collections (sets) are more widely used in computer
science in the from of approximation lattices (works by D.S. Scott
commenced in 1969, \cite{Sco:71}) with employing infinite domain.
An attempt to formalize an idea of `information system' with
infinite domain is known as continuity principle and called
Scott-Belnap's thesis). In the early 1980's D.S.Scott resumed his
analysis of the idea of variable domains because of the nature of
lambda-calculus models and related formal systems (like
combinatory logic) having been elucidated.
     Logician H.Curry and his school independently investigated the
concepts of the `object' and the `substitution' in order to render theoretical
computer science with the trustworthy representations from the early 1960's
(in fact, even earlier).
     The results obtained by H.Curry-D.S.Scott became the mathematical
foundations of modern computer science and inspired many subsequent
works, e.g. D.Turner ([1979] applicative programming),
T.Hughes ([1982] supercombinators in programming),
P.-L. Curien et.al. ([1985] categorical combinatory logic
in programming),
P. Fradet et.al. ([1990-91] functional
programming languages reducible to collections (systems) of combinators).
     All these investigations as well as adjoining contemporary works
employ the idea of `object' borrowed from the combinatory logic (in the
sense of Curry-Scott).
     The analysis of problems arose and, in particular, representation of
transition processes in information systems (dependency of program code on
the basis of combinators, transition from some basis to another etc.) reflects
the principle effect of varying the primary established objects. The attempts
of partial solutions do not result in an adequate mathematical apparatus.
Some advance was done within the {\em applicative} computational
framework (see references in~\cite{Wo:96}).

\section{Applied research}

The starting point for ADE project originates from the
various object notions.

Known results in the field of relational data bases (originated from works by
E. Codd and his school [1972 and later]), in the field of conceptual
modelling (M.Brodie [1984 and later]), in the field of knowledge based
systems (including inference and transformation mechanisms) chiefly deal
with established modes of information systems.
     Establishing the representation of a concept as invariant on variations,
creating the tools of `evolving the concept' as a mathematical process
overtakes these essential difficulties and are put into fundamentals of analysis
an information system dynamics (also the transition or switching processes).
     Developing the theory of variable concepts naturally arises from the
very process of creating theoretical foundations of computer science.
Information systems contain besides database (DB) also the metadata base
(MDB) that does include different facts concerning database. In applications
metadata bases are used as the components of knowledge system with
knowledge bases. Both parts, DB and MDB communicate each other and the
host computational environment of information system.
     The relative ideas were understood from 1987 in the nine known research
projects of {\em extensible} database systems.
De Witt ([1988] and later) gave a
brief comparison and further directions of the possible research. The main
ideas were concentrated on a model of data. Even the project without any
pre-defined data model was proclaimed, but the desirable degree of
flexibility was not achieved. The most prominent experiments with
experimental (object-oriented) extensible system have used the molecular
chaining of objects to avoid the known difficulties in designing and
development. But the completed solution was not obtained neither in a theory
nor in practice.
     The design of the systems which accumulate and process both
information and metainformation does influence the effectiveness
of information system, the efforts of its designing, maintenance,
modifications and extensions. This results in the problem of
designing the Application Development Environment (ADE) which
communicates with the varying DB and MDB.

\section{Objectives of the proposed ADE conception}

ADE is both the research project and tool kit aimed at development of
the mathematical apparatus, models and methods which being considered
together comprise a variant of theory of computations for object-oriented
programming technology, data base management systems and knowledge
based systems.
The research aims were establishing and experimental issues
of some generic concepts in computer science: data object to represent the
storable data; metadata object to represent the conceptual information;
assignment to represent internal states and dynamics; extensible database to
represent relatively self-contained couples of objects and some additional
derived concepts.
     All of these notions are important to succeed in ADE development.
     At the present time in computer science there are proper means and
methods to handle static application domains. And there is an appropriate
mathematical apparatus as well. But the work with dynamic application
domain is still a problem unsolved also in mathematical aspect. Other aspect
of the proposed research is to verify the feasibility of previously aimed
generic concepts. Thus the implementation of a prototype information system
will assist in understanding the dynamics of data objects both within tool and
applied (programming of information) system.

\section{ADE contribution to the research activity}

ADE has under research the idea of a concept as the variable entity to
possess the creation of the variable concepts and associated transition effects.
In their turn the variable concepts lead to parameterized type system. The
approach developed in ADE is based on the reasons stated.
     The usage of the method of embedding typed system (including the
apparatus of variable concepts) into untyped system based on the apparatus
of Applicative Computational Systems (ACS) is the distinctive feature of the
approach being developed.
     Combining the ideas of variable concepts will make possible
development of a wide range of applied information systems, particularly in
the field of data base management systems, knowledge based systems and
programming systems.

\section{General features}

ADE is viewed to be a comprehensive research as follows:

establishing the primitive frame to represent and analyze a `variable
concept';

setting up the approach to integrate the far distant concepts, means and
models from computer science branches;

developing methods to adopt some intentional concepts of computer
science that naturally result from the idea of `variable concepts' and their
potential applications;

creating the tool kit to explicate and apply the advantages of variable
concepts;

augmenting the possibilities of host programming system;

specifying the enhanced data models;

fixing the possible ranges of design and development those systems that
involve the idea of data/metadata object;

creating the generalized tool kit on the basis of the mathematical concepts
in the current contribution.

\begin{figure*}
\footnotesize
\unitlength=1.00mm \special{em:linewidth 0.4pt}
\linethickness{0.4pt}
\begin{picture}(150.00,156.00)
\put(60.00,145.00){\framebox(40.00,11.00)[cc]{{\bf GUI}}}
\put(30.00,90.00){\framebox(40.00,11.00)[cc]{{\bf PO} System}}
\put(115.00,91.00){\framebox(35.00,15.00)[cc]{{\bf DPO}
Interface}} \put(37.00,69.00){\framebox(40.00,11.00)[cc]{{\bf
S}torage {\bf I}nterface}}
\put(37.00,45.00){\framebox(25.00,15.00)[cc]{{\bf PO} Language}}
\put(82.00,45.00){\framebox(25.00,15.00)[cc]{{\bf PO} Relations}}
\put(51.00,28.00){\framebox(25.00,10.00)[cc]{{\bf AO} Language}}
\put(37.00,5.00){\framebox(25.00,11.00)[cc]{{\bf R}-level}}
\put(82.00,5.00){\framebox(25.00,11.00)[cc]{{\bf S}-level}}
\put(32.00,3.00){\framebox(80.00,15.00)[cc]{}}
\put(25.00,111.00){\framebox(110.00,27.00)[cb]{{\bf C}onceptual
{\bf S}hell {\bf CS}: {\bf POL} Interface}}
\put(32.00,43.00){\framebox(80.00,20.00)[cc]{}}
\put(25.00,87.00){\framebox(75.00,20.00)[rt]{{\bf C}onceptual {\bf
R}elational {\bf I}nterface}}
\put(47.00,26.00){\framebox(60.00,14.00)[rc]{{\bf R}elational {\bf
S}ystem}} \put(25.00,1.00){\framebox(90.00,83.00)[rt]{{\bf DPO}
System}} \put(30.00,124.00){\framebox(40.00,11.00)[cc]{{\bf
C}onceptual {\bf I}nterface-{\bf 1}}}
\put(85.00,124.00){\framebox(40.00,11.00)[cc]{{\bf C}onceptual
{\bf I}nterface-{\bf n}}} \put(60.00,135.00){\vector(1,1){10.00}}
\put(85.00,145.00){\vector(1,-1){10.00}}
\put(70.00,132.00){\vector(1,0){15.00}}
\put(85.00,127.00){\vector(-1,0){15.00}}
\put(115.00,101.00){\vector(-1,0){15.00}}
\put(100.00,95.00){\vector(1,0){15.00}}
\put(55.00,90.00){\vector(0,-1){6.00}}
\put(65.00,84.00){\vector(0,1){6.00}}
\put(100.00,135.00){\vector(-1,1){10.00}}
\put(65.00,145.00){\vector(-1,-1){10.00}}
\put(45.00,69.00){\vector(-1,-2){4.67}}
\put(50.00,60.00){\vector(1,2){4.67}}
\put(62.00,55.00){\vector(1,0){20.00}}
\put(82.00,50.00){\vector(-1,0){20.00}}
\put(62.00,13.00){\vector(1,0){20.00}}
\put(82.00,8.00){\vector(-1,0){20.00}}
\put(35.00,111.00){\vector(0,-1){10.00}}
\put(45.00,101.00){\vector(0,1){10.00}}
\put(125.00,60.00){\framebox(25.00,15.00)[cc]{{\bf S}earch {\bf
S}ystem}} \put(125.00,70.00){\vector(-1,0){10.00}}
\put(115.00,65.00){\vector(1,0){10.00}}
\put(29.00,25.00){\dashbox{2.00}(84.00,41.00)[cc]{}}
\put(50.00,43.00){\vector(3,-1){10.00}}
\put(80.00,40.00){\vector(3,1){10.00}}
\put(60.00,25.00){\vector(0,-1){7.00}}
\put(84.00,18.00){\vector(0,1){7.00}}
\put(65.00,69.00){\vector(3,-1){27.00}}
\put(101.00,60.00){\vector(-3,1){27.00}}
\end{picture}
\caption{ {\bf A}pplication {\bf D}evelopment {\bf E}nvironment:
{\bf ADE}. {\small\em     Abbreviations:
         {\bf GUI} - {\bf G}raphical {\bf U}ser {\bf I}nterface;
         {\bf POL} - {\bf P}otential {\bf O}bject {\bf L}ibrary;
         {\bf PO}  - {\bf P}otential {\bf O}bject;
         {\bf AOL} - {\bf A}ctual {\bf O}bject {\bf L}ibrary;
         {\bf AO}  - {\bf A}ctual {\bf O}bject;
         {\bf DPO} - {\bf D}escendent {\bf P}otential {\bf O}bject;
         {\bf R}-level - {\bf R}epresentation level;
         {\bf S}-level - {\bf S}torage level
}
}\label{pic:B}
\end{figure*}
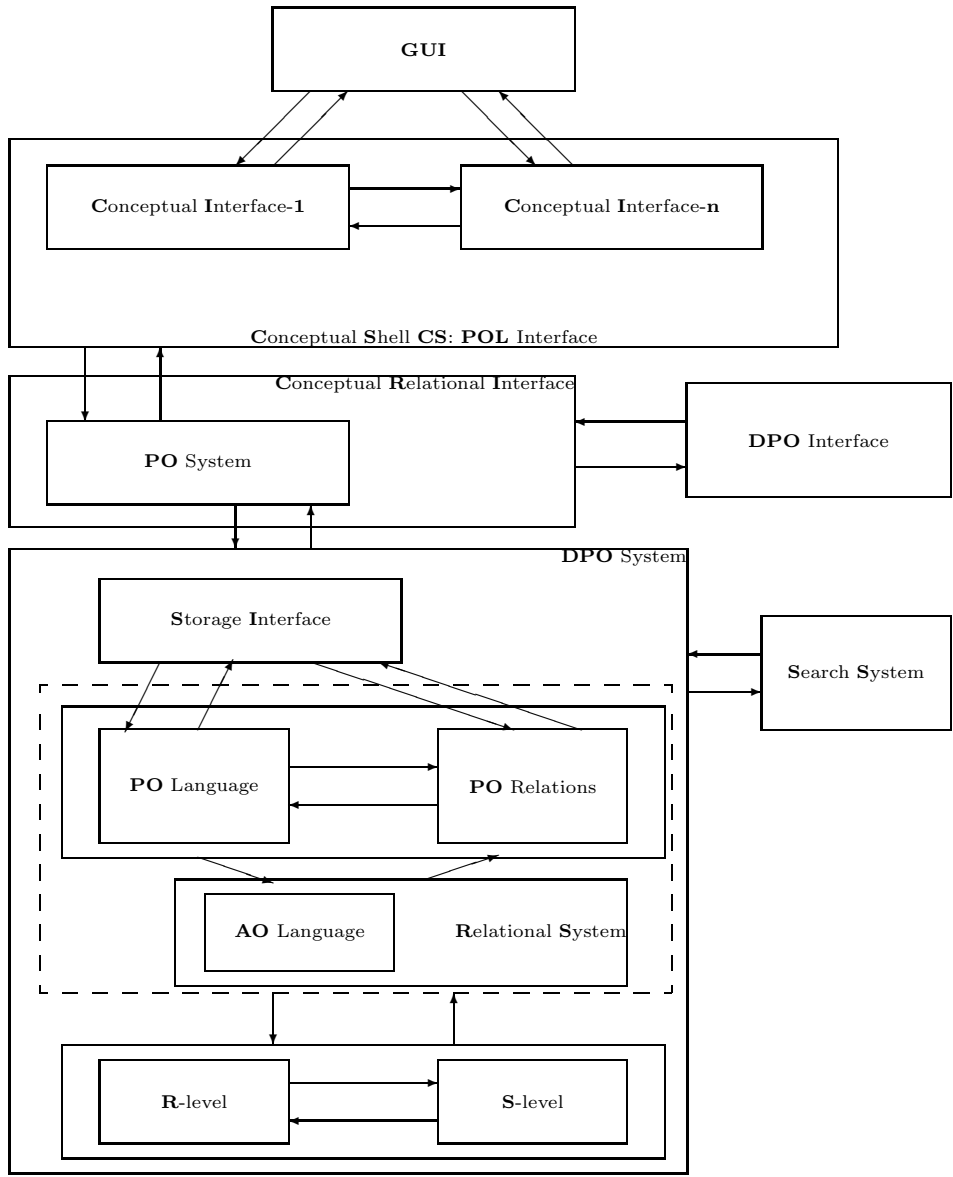

The target prototype system Application Development Environment
(ADE) is mainly based on the idea of variable, or switching
concept and covers the vital mechanisms of encapsulation,
inheritance and polymorphism. Variable concepts naturally generate
families of similar types that are derived from the generic types.
Concepts in ADE are equipped with the evolvents that manage the
transitions, or switching between the types. In particular, the
identity evolvent supports the constant concepts and types
(statical concepts). To achieve the needed flexibility a general
ADE layout consists of the uniform modular units, as shown in
Fig.~\ref{pic:B}.

In ADE Data Object Definition Language (DODL) contains the
construction of data objects' base scheme as a relation between
concepts. Concepts are included into the type system with the
interpretation over the variable domains. A coherent set of
variable domains generates the data objects' base. Basis to
maintain the data objects in use and their bases is generated by
computational models with applicative structures. The developer
obtains the set of the means that establish, support and modify
the linkages between the data objects' base schemes, data objects'
base and computational models. DODL declares: {\em type system} as
a set of metadata objects; em linkages between the types; system
of domains; linkages between the domains; extensions of domains
and types; computational tools of applicative pre-structures and
structures.

The third part of the implementation supports two level of
interfaces. The first is the Intentional Management System (IMS)
to support concepts (metadata objects) of different kinds, and the
second is associated Extensional Management System (EMS) to
support the appropriated extensions (data objects) generated by
the intentions.

EMS is embedded into the unified computational model. It is
object-oriented extensible programming system Basic Relational
Tool System (BRTS). BRTS has the fixed architecture with the one
level comprehension, separate self- contained components,
interfaces and languages. It is the First Order Tool (FOT) and
generates `fast prototypes'. D(M)ODL and D(M)OML of BRTS contain
the SQL-based relational complete languages that cooperate with
ADE. BRTS mainly supports relatively large number of low
cardinality relations (extensions) and supports Data (Metadata)
Object Model D(M)OM with retrieval, modifications and definitions
of a storable information.

IMS is also embedded into the computational model and supports a numerous
matadata objects. Their amount is almost the same as for data objects. IMS
is based on D(M)OM with a simple comprehension to manage metadata base
and is supported by MetaRelational Tool System (MRTS). MRTS
manipulates with the metaobjects (concepts) and metarelations (frames) and
is embedded into ADE.

\section{New supporting technologies}

A main result is the experimental verification of variable concepts approach.
This would be applied to develop the variety of applied information systems.

Computations with variable concepts and appropriated programming
system allows to built a system especially to manipulate the
objects. All of this results in an apparatus for analysis the
branches of object-oriented approach:

programming in terms of concepts. This generalizes the programming in
terms of superobjects;

object oriented database (management) systems. The database
extensibility, development of databases with varying data models,
dynamical databases, analyzers of `data object' dynamics, the
`bases of invariants' etc. would be achieved;

knowledge base systems grounded on the concepts which are metadata
objects. Maintenance of the systems of `variable concepts' and their
interrelations, management of switching the systems of concepts,
management of database modification etc.

\section{Usage of intentions and extensions}

ADE gives a smart framework for intentions and extensions and
intentional tool/applied system. It enables the possibility to
develop the conceptual support that encircles the adjunct ideas:

establishing the logical apparatus (on the basis of higher order theory) to
study the hierarchies of variable concepts;

development of specialized `tool theories' to estimate a selective power of
newly designed programming systems, database systems and models for
systems with databases and knowledge bases;

support and development of specialized semantical theories and models for
systems with databases and knowledge bases.

Possible applications of this framework: `rapid prototypes' of
newly developed computer information systems, estimation of their
ranges, adaptation to variations in problem domains, demands,
programming systems, experts. The specific feature of
architecture: two levelled design - intentional and extensional
levels.

\section{Improvements in design, engineering and management processes}

At present it is difficult to estimate real benefits of using the apparatus of
variable concepts in particular applied information systems because of the
newness and the originality of mathematical apparatus being developed and
the approach on the whole. The possible gains and prospects that may be
yielded by the ADE approach are set forth below:

high degree of generalization;
taking into account the intentional features (`subjects',
dynamical scripts', etc.);
relative simplicity of interfaces;
possibility to aggregate/disaggregate the representations;
clarity, referential transparency and fully explicit constructions;
possibilities to handle collections of objects or concepts;
the modular implementation to cover the higher order logic.

\section{Experimental techniques, Software, and Tools}

The feature of current research is in primary
creation of the needed tool system to be adequate to the newly generated
mathematical apparatus.
The experimental research and verification of the obtained
model is based on prototypes - CS, ADE, BRTS, BMRS. The difficulties to
implement full scale prototype are resolved by the high level object-oriented
programming language. Some candidate programming systems are tested to
enable the needed computational properties. After that the main
programming tool kit is selected. Preliminary candidate tools were C++ or
Modula-2. An attention is paid to select an appropriate database management
system. If needed the original DBMS is attached. At the preliminary tests the
attention was paid to OLE-2 techniques.

Some ready made original systems were tested and expanded to
achieve the prototype system with the properties
mentioned (by E.Codd, by N.Roussoupolos).

\section*{Conclusions: interpretation of the results}
\addcontentsline{toc}{section}{Conclusions}

The resulting two level comprehension model and computational environment
verify the feasibility of the approach. The adequate, neutral and semantical
representation of data is the target in the sphere of extensible systems and
their moderations and modifications.
The relational solutions are the criteria in database technology.
Therefore, the
variable concepts generate the power and sound representation of data
objects, have the boundary conditions as the known results in information
systems (both in a theory and applications) and capture the additional effects
of dynamics to simulate, in particular, the encapsulation, polymorphism and
inheritance. The last gives the contribution in development of object-oriented
systems.


\addcontentsline{toc}{section}{References}
\newcommand{\noopsort}[1]{} \newcommand{\printfirst}[2]{#1}
  \newcommand{\singleletter}[1]{#1} \newcommand{\switchargs}[2]{#2#1}


\end{document}